\documentclass[runningheads]{llncs}
\usepackage{graphicx}
\usepackage{booktabs} 
\usepackage[T1]{fontenc}
\usepackage{listings}
\usepackage{color}
\definecolor{bluekeywords}{rgb}{0.13,0.13,1}
\definecolor{greencomments}{rgb}{0,0.5,0}
\definecolor{greynumbers}{rgb}{0.46,0.45,0.48}
\definecolor{redstrings}{rgb}{0.5,0,0}

\lstdefinelanguage{astra}{
    keywords = {agent, module, extends, inference, rule, if, for, while},
    morestring=[b]",
    basicstyle=\scriptsize\ttfamily,
    keywordstyle=\scriptsize\color{bluekeywords},
    stringstyle=\scriptsize\color{redstrings},
    upquote=true
}

\begin{document}
\title{Using Multi-Agent MicroServices (MAMS) for Agent Based Modelling}
\author{Martynas Jagutis \and Sean Russell \and Rem W. Collier}
%
%
\institute{School of Computer Science, University College Dublin, Dublin, Ireland
\email{\{rem.collier,sean.russell\}@ucd.ie}}

\maketitle
\begin{abstract}
This paper demonstrates the use of the Multi-Agent MicroServices (MAMS) architectural style through a case study based around the development of a prototype traffic simulation in which agents model a population of individuals who travel from home to work and vice versa by car.

\end{abstract}

\section{Introduction}
Multi-Agent MicroServices (MAMS) \cite{w2019mams} is an architectural style for deploying Multi-Agent Systems (MAS) within Microservices architecture. This has been achieved by introducing of a specific kind of agent, known as a \textit{MAMS Agent}, that has an associated body that consists of a set of web resources that are accessible through REpresentational State Transfer (REST). MAMS agents act like an interface agent for a microservice. Collectively, their bodies form a REST interface that external microservices can use to interact with the MAS via the MAMS agents. Within a microservice, MAMS agents are able to interact with non-MAMS agents through traditional agent communication mechanisms.

MAMS has been applied to a number of problem domains, including: decision support tools \cite{carneiro2020consensus}, building management \cite{o2021building} and digital twins for smart agriculture \cite{kalyani2023towards}.  Additionally, a prototype framework for implementing MAMS applications \cite{o2020delivering}  has been developed, built on a combination of CArtAgO \cite{ricci2006cartago} and the ASTRA programming language \cite{collier2015reflecting}. The source code for the framework and a number of example applications can be found on Gitlab\footnote{\url{https://gitlab.com/mams-ucd/}}.

This paper illustrates a potential use of MAMS and microservices in Agent Based Modelling (ABM) \cite{abar2017agent}. The basic idea is to decompose the environment part of an ABM into a set of web resources. For example, a road network can be decomposed into street and junction resources. Each resource is a kind of \textit{"micro-environment"} that agents can inhabit and interact with. They are created and accessed through specially designed environment microservices. Inter-resource relationships are modelled based on the URL associated with each resource. A second set of microservices are used to implement the agent part of the ABM by leveraging the MAMS architectural style.

\section{Overview of Prototype}\label{sec:mts}
The scenario demonstrated in this paper is a simple traffic simulation scenario in which agents model a population of individuals who travel from home to work and vice versa by car.
The environment for this scenario is decomposed into four types of resource: home resources, work resources and the street and junction resources that model the road network.
These resources are implemented through three \textit{sub-environment} microservices.
The design of the street and junction resources is based on best practices drawn from established traffic simulators such as MATSim \cite{w2016multi} and SUMO \cite{lopez2018microscopic}. 

\begin{figure}[h!]
	\includegraphics[width=\textwidth]{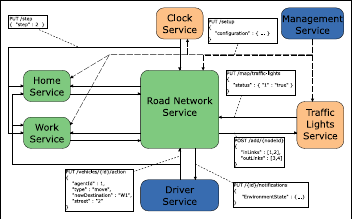}
	\caption{Overview of Simulation Architecture}
	\label{fig:1}
\end{figure}

Figure~\ref{fig:1} illustrates the set of microservices, implemented using \textit{Java} and \textit{Spring Boot}\footnote{\url{https://spring.io}}, that underpin the prototype. This includes three sub-environment microservices described above. The \textbf{Road Network} service is the most complex of the three and is underpinned by the \textit{Neo4J database}\footnote{\url{https://neo4j.com/}} which maintains a graph of the constituent streets and junctions. The \textbf{Home} and \textbf{Work} services provide a minimal model that includes access to the current time and a single activity (e.g. \textit{Watch TV} or \textit{Work}).
A \textbf{Clock Service} provides a discrete time model for the simulation; a \textbf{Traffic Lights Service} implements an algorithm to control traffic lights in the \textbf{Road Network} and a \textbf{Management Service} supports the configuration and execution of a simulation run.
Finally, the \textbf{Driver Service} implements the agent part of the system which is described next.

To connect to the simulation, MAMS Agents must register with an environment microservice based on the resource they wish to interact with. The microservice can reject the request, but if accepted, it creates an agent body resource\footnote{This is not the same as the MAMS body described above}. The environment state is passed to the agent using a HTTP PUT Request to a webhook associated with the MAMS agents body. Agents submit actions using a HTTP PUT Request to the agents body on the environment microservice. The environment microservice tracks which resource an agent is associated with. When an agent moves to another resource (e.g. moving from a junction to a street), the microservice registers the change. If the agent moves to a resource that is located on a different microservice, its body is transferred to the new microservice via a HTTP POST Request. Further details can be found in \cite{idc2022} and the source code is available on Gitlab\footnote{\url{https://gitlab.com/mams-ucd/examples/microservice\_traffic\_simulator}}.

\section{Using MAMS to implement agent behaviours}\label{sec:mamsdriver}
This section focuses on the implementation of the \textbf{Driver Service} using the MAMS prototype that has been developed for the ASTRA agent programming language. A simplified version of the driver agent implementation used in the demo is shown in Figure~\ref{fig:astradriver}. The overall behaviour begins with the handling of the \verb|!updatedObject(...)| goal in the second rule.  The argument of this goal is a Java object that represents the environment state. The plan part of the plan rule defines two sub-goals \verb|!decide(...)| and \verb|!act(...)| which must be achieved in sequence. The last two rules in the program highlight two possible sub-plans for achieving the \verb|!decide(...)| goal. The agent will choose only one of these options based on the current state of the environment. For example, the last rule requires that the vehicle controlled by the agent be stopped. This is expressed by the \verb|isStopped(...)| belief.  This belief is evaluated based on the second of the inference rules at the top of the code snippet, which are denoted by the \verb|inference| keyword.  The \verb|ObjectAccess| module provides a generic mechanism for the agent to query the internal state of Java objects. In this case it retrieves the value of the \verb|vehicleSpeed| field of the \verb|EnvironmentState| object. The selection conditions are expressed by the context part which appears after the colon (:) and before the opening brace (\{) of the plan. The \verb|!act(..)| goal sends the chosen action to the server using the low level \verb|!put(...)| goal provided by the MAMS implementation.

\begin{figure}[t]
    \begin{lstlisting}[language=astra,basicstyle=\scriptsize,frame=ltrb]
agent Driver extends mams.PassiveMAMSAgent {
  module ObjectAccess oa;

  inference atIntersection(EnvironmentState state) :- 
      oa.isFalse(state, "atIntersection") &
      oa.getInt(state, "vehicleSpeed") > 0;
      
  inference isStopped(EnvironmentState state) :- 
      oa.getInt(state, "vehicleSpeed") > 0;

  rule +!main(list args) {
     MAMSAgent::!init();
     MAMSAgent::!created("base");
     PassiveMAMSAgent::
        !itemResource("notifications", "EnvironmentState");
  }
  
  rule +!updatedObject(EnvironmentState state) {
    !decide(state, oa.getString(state, "type"), string action);
    !act(state, action);
  }
  
  rule +!act(EnvironmentState state, string action) {
    !put(oa.valueAsString(state, "webhook"),
        "{ 'action':'"+action"'}", HttpResponse response);
    if (!httpUtils.hasCode(response, 200)) system.fail();
  }
  
  rule +!decide(EnvironmentState state, "traffic", string action)
        : time(t) & atIntersection(state) {
    action = "move";
  }
  
  rule +!decide(EnvironmentState state, "traffic", string action)
        : time(t) & isStopped(state) & canAccelerate(state) {
    action = "accelerate";
  }
}
    \end{lstlisting}
    \caption{ASTRA-MAMS Implementation}
    \label{fig:astradriver}
\end{figure}

MAMS is visible in two parts of the example code. The latter place is in the rule associated with the \verb|!act(...)| goal where the \verb|!put(...)| goal is adopted to submit the action to the server. The representation actually sent to the simulation has been simplified for readability. The former place where MAMS is visible is in the rule that handles the \verb|!main(...)| goal. The goals specified in this rule connect the agent to the MAMS infrastructure and create a resource that is exposed on the web under the \verb|/{agent-name}/notification| URL.  The simulation service sends the environment state to the agent in the same way; by updating this resource using a PUT request. Upon the processing of a new PUT request, the underlying MAMS infrastructure generates the \verb|updatedObject(...)| goal to trigger a response from the agent.

\section{Conclusions}
This paper presents an early prototype of an novel approach to Agent Based Modelling (ABM) using a combination of microservices and the Multi-Agent MicroServices (MAMS) architectural style. The prototype presented is a traffic simulation scenario that decomposes the environment into four types of web resource that are hosted across three microservices. Each resource acts as a "micro-environment". Agents interact with a resource by registering a "body" with the corresponding microservice, indicating which resource they wish to be associated with. Hypermedia links are used to relate resources to one another, for example, a junction resource in the road network can be linked to a home or work resource. A key part of the approach is the design of mechanisms to allow agents to transition between web resources which can be achieved either internally or via a HTTP POST request. 

A number of shortcomings and opportunities were identified during its evaluation \cite{idc2022}. The most interesting opportunity is the potential use of the linked data structure to create decentralised knowledge graphs that capture global knowledge of the simulation environment. Such knowledge could be consumed by individual agents and used in concert with local contextual knowledge of their environment to offer improved decision making capabilities. Details of this proposed approach can be found in \cite{collier2022towards}. 

%
%
%

\bibliographystyle{splncs04}
\bibliography{mybibliography}

\begin{thebibliography}{10}
\providecommand{\url}[1]{\texttt{#1}}
\providecommand{\urlprefix}{URL }
\providecommand{\doi}[1]{https://doi.org/#1}

\bibitem{abar2017agent}
Abar, S., Theodoropoulos, G.K., Lemarinier, P., O’Hare, G.M.: Agent based
  modelling and simulation tools: A review of the state-of-art software.
  Computer Science Review  \textbf{24},  13--33 (2017)

\bibitem{carneiro2020consensus}
Carneiro, J., Andrade, R., Alves, P., Concei{\c{c}}{\~a}o, L., Novais, P.,
  Marreiros, G.: A consensus-based group decision support system using a
  multi-agent microservices approach. In: Proceedings of the 19th International
  Conference on Autonomous Agents and MultiAgent Systems. pp. 2098--2100 (2020)

\bibitem{w2019mams}
Collier, R., O'Neill, E., Lillis, D., O'Hare, G.: Mams: Multi-agent
  microservices. In: Companion Proceedings of The 2019 World Wide Web
  Conference. pp. 655--662. ACM (2019)

\bibitem{collier2022towards}
Collier, R., Russell, S., Ghanadbashi, S., Golpayegani, F.: Towards the use of
  hypermedia mas and microservices for web scale agent-based simulation. SN
  Computer Science  \textbf{3}(6), ~510 (2022)

\bibitem{collier2015reflecting}
Collier, R.W., Russell, S., Lillis, D.: Reflecting on agent programming with
  agentspeak (l). In: International Conference on Principles and Practice of
  Multi-Agent Systems. pp. 351--366. Springer (2015)

\bibitem{idc2022}
Jagutis, M., Russell, S., Collier, R.: Simulating traffic with agents,
  microservices \& rest. In: 15th International Symposium on Intelligent
  Distributed Computing (IDC2023) (2022)

\bibitem{kalyani2023towards}
Kalyani, Y., Collier, R.: Towards a new architecture: Multi-agent based
  cloud-fog-edge computing and digital twin for smart agriculture. In:
  Intelligent Distributed Computing XV, pp. 111--117. Springer (2023)

\bibitem{lopez2018microscopic}
Lopez, P.A., Behrisch, M., Bieker-Walz, L., Erdmann, J., Fl{\"o}tter{\"o}d,
  Y.P., Hilbrich, R., L{\"u}cken, L., Rummel, J., Wagner, P., Wie{\ss}ner, E.:
  Microscopic traffic simulation using sumo. In: 2018 21st international
  conference on intelligent transportation systems (ITSC). pp. 2575--2582. IEEE
  (2018)

\bibitem{o2021building}
O’Neill, E., Beaumont, K., Bermeo, N.V., Collier, R.: Building management
  using the semantic web and hypermedia agents  (2021)

\bibitem{o2020delivering}
O’Neill, E., Lillis, D., O’Hare, G., W~Collier, R.: Delivering multi-agent
  microservices using cartago. In: International Workshop on Engineering
  Multi-Agent Systems. pp. 1--20. Springer (2020)

\bibitem{ricci2006cartago}
Ricci, A., Viroli, M., Omicini, A.: Cartago: A framework for prototyping
  artifact-based environments in mas. In: International Workshop on
  Environments for Multi-Agent Systems. pp. 67--86. Springer (2006)

\bibitem{w2016multi}
W~Axhausen, K., Horni, A., Nagel, K.: The multi-agent transport simulation
  MATSim. Ubiquity Press (2016)

\end{thebibliography}
\end{document}